\documentclass[aps,nofootinbib,preprintnumbers,citeautoscript]{revtex4-2}
\usepackage{amsmath,amssymb}
\usepackage{enumerate}
\usepackage{graphicx}
\usepackage{amsthm}
\usepackage{braket}
\usepackage{mathtools}
\usepackage{textcomp}
\usepackage{amssymb}
\usepackage{wasysym}
\usepackage{subcaption}
\usepackage{amsmath}
\usepackage{lineno}

\usepackage{standalone}
\usepackage{url}

\begin{document}

		\title{Variational Quantum Classifiers Through the Lens of the Hessian}
	
	    \author{ Pinaki Sen$^{1}$, Amandeep Singh Bhatia$^{2^*}$, Kamalpreet Singh Bhangu$^{3}$, Ahmed Elbeltagi$^{4}$  \\
		\textit{$^{1}$National Institute of Technology, Agartala, Tripura, India} \\
		\textit{$^{2,3}$Chitkara University Institute of Engineering \& Technology, Chitkara University, Punjab, India} \textit{$^{4}$Agricultural Engineering Dept., Faculty of Agriculture, Mansoura University, Mansoura 35516, Egypt}\\
		E-mail: $^{*}$amandeepbhatia.singh@gmail.com}

	\begin{abstract}
	In quantum computing, the variational quantum algorithms (VQAs) are well suited for finding optimal combinations of things in specific applications ranging from chemistry all the way to finance. The training of VQAs with gradient descent optimization algorithm has shown a good convergence. At an early stage, the simulation of variational quantum circuits on noisy intermediate-scale quantum (NISQ) devices suffers from noisy outputs. Just like classical deep learning, it also suffers from vanishing gradient problems. It is a realistic goal to study the topology of loss landscape, to visualize the curvature information and trainability of these circuits in the existence of vanishing gradients. In this paper, we calculate the Hessian and visualize the loss landscape of variational quantum classifiers at different points in parameter space.  The curvature information of variational quantum classifiers (VQC) is interpreted and the loss function's convergence is shown. It helps us better understand the behavior of variational quantum circuits to tackle optimization problems efficiently. We investigated the variational quantum classifiers via Hessian on quantum computers, starting with a simple 4-bit parity problem to gain insight into the practical behavior of Hessian, then thoroughly analyzed the behavior of Hessian's eigenvalues on training the variational quantum classifier for the Diabetes dataset. Finally, we show how the adaptive Hessian learning rate can influence the convergence while training the variational circuits.
 \\ \\

	\textbf{Keywords}: Quantum machine learning, quantum algorithms, classification, loss landscape, variational quantum circuit.

	\end{abstract}
	
	\pacs{Valid PACS appear here}
	\maketitle
	

\section{Introduction \& Motivation}
In recent years, the enhancement of machine learning algorithms by noisy intermediate-scale quantum (NISQ) technology and mainly the variational quantum circuits have garnered significant attention among academic and research communities \cite{1}.  Researchers have applied the variational quantum algorithms in various applications in the NISQ era, mainly the ones related to quantum artificial intelligence. The variational quantum algorithms (VQAs) have shown great learning capability to counterbalance the errors in the device framework. They are considered to be the greatest hope for the journey toward quantum advantage. The first variational quantum eigensolver (VQE) was proposed as a state ansatz to determine the ground state energy of physical systems \cite{2}. Since the first VQE introduced, several VQE variants have been proposed with a plethora of alterations for computation of excited states such as orthogonality constrained VQE \cite{3}, subspace approach VQE \cite{4, 5}, adiabatically assisted VQE \cite{6} and multistate contracted VQE \cite{7}. These variational quantum circuits are constructed with an ansatz whose parameters are trained with several optimization methods to minimize the cost. For combinatorial optimization tasks, the quantum approximate optimization algorithm (QAOA) was proposed originally to attain approximate solutions \cite{8}. These architectures of variational quantum circuits have shown to be computationally universal \cite{9, 891}.

Recently, the use of VQAs has got an incredible response and is widely applied in quantum machine learning applications. VQAs consist of a small number of qubits and quantum circuits, which make them resistant to noise. The VQAs are effective for a classification task in machine learning; the objective is to 
train a classifier and predict the label of each input accurately \cite{10, 88, 890}.  Suppose, a training data is given $\{x_i, y_i\}$, where $x_i'$s and $y_i'$s are inputs and labels respectively. The variational quantum circuit is used as a black box to predict the right label ($y_i$) for each input after embedding the classical data into the quantum states. VQAs are the quantum variants of neural networks, the most commonly used and highly successful machine learning model. Till now, several architectures for quantum neural networks (QNNs) have been proposed and applied in different areas \cite{11, 12, 13, 14}. Recently, Cong et al. \cite{15} introduced the quantum convolutional neural networks (QCNNs) and used them to discriminate quantum states of distinct topological phases. Bhatia et al. \cite{78} performed the simulation of several entangled states on a quantum computer.  Romero et al. \cite{16} proposed a quantum variational autoencoder for compressing the quantum data efficiently. Pepper et al. \cite{17} demonstrated the experimental realization of a proposed quantum autoencoder, which will likely be an essential primordial in quantum machine learning. In recent years, generative adversarial networks (GANs) have been an exciting topic of research in classical machine learning. Romero and Aspuru-Guzik \cite{18} introduced the quantum variant of GAN for learning continuous distributions and to speedup classical GAN using quantum systems. Kappen \cite{72} proposed a method to represent the classical data distribution in a quantum system. Moreover, the potential of VQAs can be evaluated in several industry based applications such as supply chain management, intelligent healthcare, smart agriculture, manufacturing production. and cloud manufacturing \cite{810}.

\begin{figure*}[!ht]
	\includegraphics[scale=0.4]{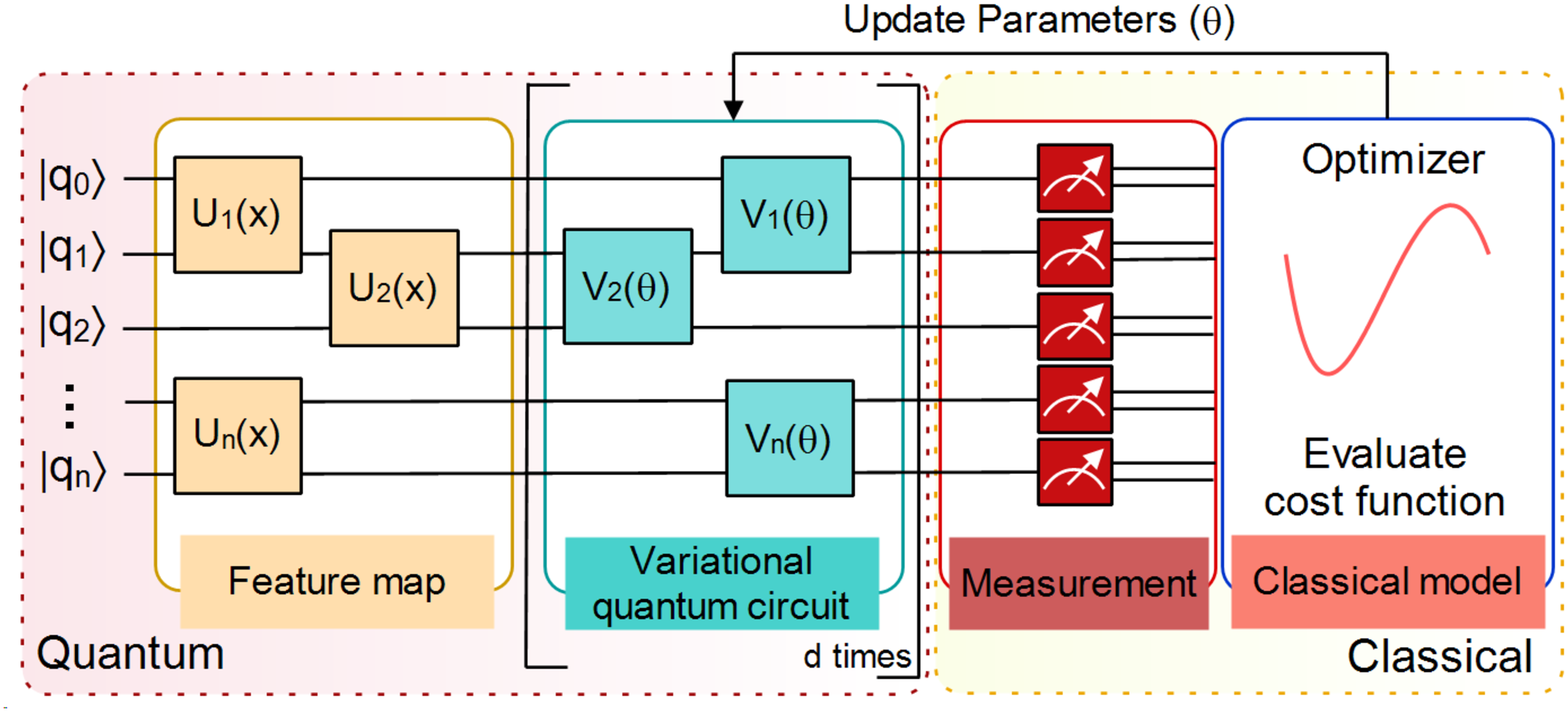}
	\caption{\textbf{Schematic representation of a variational quantum circuit (VQC).} VQC is created by combining the feature map, variational circuit and measurement component. It is followed by classical optimization model. It includes the iterative implementation of quantum and classical components of the circuit.}
\end{figure*}

The major downside of variational quantum circuits is the occurrence of a barren plateau, which vanishes the gradients of cost function exponentially with the increase in the number of qubits \cite{45, 46}. It can be represented by flat plateaus of the loss function \cite{48}. Firstly, in 2018,  McClean et al. \cite{19} studied the barren plateau phenomenon numerically for a class of random quantum circuits. It has been shown that variance of the gradients of modest size quantum circuits vanishes exponentially. Cerezo et al. \cite{20} described a barren plateau phenomenon as a point in the parameter space and investigated it for shallow quantum neural networks (QNNs). It has been proven that global cost gradients vanish exponentially at all depths, whereas local costs exhibit non-vanishing gradients at shorter depth quantum circuits. Grant et al. \cite{21} investigated the problem of a barren plateau of parametrized quantum circuits in the energy landscape. The approach is based on the initialization of parameters to avoid the barren plateaus. Initially, some parameter values are selected randomly, and the remaining values are chosen so that it becomes a chain of shallow circuits. The depth of circuits is used to determine the first updated parameter not to be lodged in a barren plateaus problem during training. Recently, it has investigated that the barren plateau is missing in QNNs and QCNNs with tree tensor network (TTN) architecture \cite{22, 23}.

In classical machine learning, the loss landscapes of neural networks and their characterization via the lens of Hessian have been well investigated. Several works have explored the flatness of local minima using the eigenvalues of Hessian \cite{23, 24, 25, 26, 32}. During training, the generalization ability of the neural networks depends on batch size. It has explored that mini-batch size tends to favor flat minima basins of the loss function with several eigenvalues of the Hessian $\lambda_j=0$ and hardly any $\lambda_j < 0$ \cite{27}. Perez-Salinas et al. attained fast convergence for some quantum classification problems using Hessian based optimization method called batched optimization. Rebentrost et al. \cite{28} proposed the quantum variants of two popular iterative optimization algorithms, Newton's and gradient descent. At time \textit{t}, numerous copies of quantum state $\ket{\psi}_{t}$ are used to generate other numerous copies of quantum state $\ket{\psi}_{t^{'}}$ at $t'=t$+1 by using the Hessian of an objective function and in consideration of a gradient vector. Recently, Huembeli and Dauphin \cite{29} calculated the Hessian of a loss function and characterized the loss landscape of variational quantum circuits. It showed that the Hessian helps escape the flat surface of loss function for certain data-driven variational quantum circuits.

The VQCs have been extensively employed in a wide array of new applications. Fig 1 shows the schematic representation of a variational quantum circuit. It involves evaluating a cost function or its gradient on a quantum system \cite{41, 47}. A classical optimization loop trains the parameters ($\theta$) of a variational quantum circuit $V(\theta)$ to reduce the cost.
It is well-known that the feature map encodings in variational parameterized circuit architecture designs produce loss functions that train easier, and well-selected training parameters "\textit{optimizer}" that generalize well. Nevertheless, the effects of parameters on the entire loss landscape are not well studied. It has not received the attention it deserves.  The occurrence of barren plateaus issue could abolish the quantum advantage with a parameterized quantum circuit \cite{42}. The visualization and understanding of the loss landscape of classical machine learning algorithms remains a vital and highly active area of research. Due to issues of computational complexity, the structure of loss landscape of variational quantum classifiers is not well visualized and recognized as compared to classical neural networks. A better visualization can really work for the advancement of optimization algorithms and can highlight the shortcomings of quantum circuit designs. Hence, it is a natural goal to study the loss landscape of variational quantum classifiers with the eigenvalues of Hessian to recognize when the quantum speedup is achievable. The following contributions are claimed:
\begin{itemize}
	\item Visualized the loss landscape of variational quantum classifiers at different points in parameter space using Hessian matrices.
	\item Analyzed the behavior of Hessian’s eigenvalues on training the variational quantum classifier for different datasets. 
	\item Investigated that how the adaptive Hessian learning rate can influence the convergence while training the variational circuits.
	\item Observed that adaptive Hessian learning rate can help to overshoot the cost if it gets stuck into local minima and converge quickly.
\end{itemize}

The eigenvectors and eigenvalues of Hessian present a clear interpretation of the loss landscape of a VQC. We started with a 4-bit parity problem to provide perception about the behavior of Hessian and then studied the VQC trained on diabetes classical data acting as a classifier. The organization of the rest of this paper is as follows: Sect. 2 is devoted to preliminaries and Hessian's computation of variational quantum classifier. In Sect. 3, we computed the Hessian on a quantum simulator and visualized the curvature information of a parity function. In Sect. 3 (B), we characterized the loss landscape (i.e., curvature information) of data-driven variational quantum classifiers via Hessian of the loss function, and the experimental results are plotted for the diabetes dataset. In Sect. 4, we show how the adaptive Hessian learning rate can help to overshoot the cost that helps it avoid getting stuck in local minima during training of variational circuits. Finally, Sect. 5 is the conclusion.

\section{Preliminaries}
In this Section, some basic concepts of loss function visualization with the Hessian matrix are given. Here, we give the background required to understand our results. Consider a real-valued function $f(\theta)=f(\theta_1,\theta_2,\dots,\theta_n$) with $\theta$=($\theta_1,\theta_2,\dots,\theta_n)$. The Hessian matrix ($H^2 f(\theta)$) of $f(\theta)$ is represented as the square matrix of the second derivatives of a real-valued function of \textit{n} variables, which help us to characterize the loss landscape (i.e. curvature information) \cite{30, 31}. The gradient ($\triangledown f(\theta)$) gives the partial derivatives of the function. The Hessian operator $H^2 f(\theta)$  gives the partial derivatives of the gradient \cite{43}. 
\begin{equation}
\triangledown f(\theta)= \begin{bmatrix}
	\dfrac{\partial f}{\partial\theta_1}\\
	\vdots\\
	\dfrac{\partial f}{\partial\theta_n}
\end{bmatrix},
H^2f(\theta)= \begin{bmatrix}
\dfrac{\partial^2 f}{\partial\theta_1^2} & \cdots & \dfrac{\partial^2 f}{\partial\theta_1 \theta_n}\\
\vdots & \ddots & \vdots \\
\dfrac{\partial^2 f}{\partial\theta_n \theta_1}& \cdots & \dfrac{\partial^2 f}{\partial\theta_n^2}
\end{bmatrix}
\end{equation}
If the second derivatives are continuous, then the Hessian is symmetric. The Hessian matrix consists of information about the geometric information of the function. Its eigenvalues ($\lambda_1, \lambda_2, \dots, \lambda_n$)'s are real and used to get this curvature information. So the only thing to examine is whether the eigenvalues $\lambda_i$'s are negative or positive. Suppose, $x_i$ is an eigenvector  associated with $\lambda_i$, then it is represented as $i^{th}$ eigenpair ($\lambda_i, x_i$) of \textit{H}. Using the Hessian matrix,  we can determine whether the certain point $\theta$ on a surface is locally positive or negative. The eigenvalues of the Hessian matrix give the directions of the derivatives. Suppose the Hessian evaluated at a given point $\theta$  consists of all positive eigenvalues $\lambda_i>0$ (i.e., positive-definite matrix). In that case, it shows a local positive curvature, and $\theta$ is a local minimum of \textit{f}. Similarly, if all the eigenvalues are negative, it shows a locally negative curvature, and $\theta$ is a local maximum of \textit{f}. The zero eigenvalues indicate the zero curvature of the function or flat directions. If the eigenvalues are mixed (some positive, some negative), then the surface has a saddle point $\theta$ of \textit{f}. Thus, the Hessian can be used to determine the convexity and concavity of a function of one or two variables \cite{33}. 

\subsection{Hessian Computation of VQC}
In classical machine learning, the neural networks are trained over a dataset consisting of feature vectors \{$x_i$\} and labels \{$y_i$\} by reducing the cost function as
\begin{equation}
	C(\theta)= \dfrac{1}{n} \sum_{j=1}^{n} c(l(\theta, x_j),y_j),
\end{equation}
where the $l(\theta, x_j)$ is the prediction parameterized by weights ($\theta$) of the neural network, $c$(.) denotes the loss function that measures how well the neural network predicts the label by calculating its difference with the neural network prediction, and \textit{n} denotes the size of data samples. The loss functions exist in a high-dimensional space due to the presence of several parameters in neural networks. Therefore, its visualization is not possible in higher-dimensional space. Analogously, the loss landscape of VQC has not been extensively examined as compared to classical neural networks. In a quantum layer, the classical data ($\overrightarrow{x}$) is encoded into the quantum state $\ket{f(\theta, \overrightarrow{x})}$ using a feature map consisting of quantum gates with parameters. In classical layer, the parameterized function  $f( \theta, \overrightarrow{x})$ is evaluated depending upon the learning parameters ($\theta$) in variational quantum circuit on performing measurement. The tuning of parameters ($\theta$) is executed by minimizing a loss function on a classical computer.

The gradient of a quantum circuit can be evaluated by estimating the expectation value of an observable with reference to $\theta$. It consists a series of unitary transformations. The gradient of an expectation value of an observable $\langle f(\theta, \overrightarrow{x})\rangle= \langle  \overrightarrow{x}|V^{+}(\theta) f V(\theta)|\overrightarrow{x}\rangle$ can be given as
\begin{equation}
\dfrac{\partial \langle f(\theta, \overrightarrow{x})\rangle }{\partial \theta_j}=\dfrac{1}{2}\Big[\Big\langle f\Big(\Big(\theta_{\overline{j}}, \theta_{j}+\dfrac{\pi}{2}\Big),\overrightarrow{x}\Big)\Big\rangle-\Big\langle f\Big(\Big(\theta_{\overline{j}}, \theta_{j}-\dfrac{\pi}{2}\Big),\overrightarrow{x}\Big)\Big\rangle\Big]
\end{equation}
where $\theta_{\overline{j}}$ denotes the all parameters except $\theta_{j}$,  \textit{V} is a product of unitary matrices, $\theta_{i}$ is an angle which parametrized the $V(\theta)=e^{-j \theta_j P_j/2}$, where $P_j$ is a Hermitian operator with eigenvalues $\pm1$. Let us now define the Hessian matrix elements. The Hessian (H) of a quantum circuit can be computed by performing the parameter shift rule two times \cite{34, 29, 35, 36}.
\begin{equation}
H_{ij}=\dfrac{\partial^2 \langle f(\theta, \overrightarrow{x}) \rangle}{\partial \theta_i \partial \theta_j}= \partial_i \partial_j \langle f(\theta, \overrightarrow{x})\rangle
\end{equation}
\begin{equation}
	\begin{split}
	H_{ij}= & \dfrac{1}{4}\Big[\Big \langle f\Big(\Big(\theta_{\overline{j}, i},  \theta_j + \dfrac{\pi}{2},   \theta_i + \dfrac{\pi}{2}\Big), \overrightarrow{x}\Big) \Big \rangle \\
	& - \Big \langle f\Big(\Big(\theta_{\overline{j}, i}, \theta_j + \dfrac{\pi}{2}, \theta_i - \dfrac{\pi}{2}\Big), \overrightarrow{x} \Big) \Big \rangle\\
   &- \Big \langle f\Big(\Big(\theta_{\overline{j}, i}, \theta_j -\dfrac{\pi}{2}, \theta_i +\dfrac{\pi}{2}\Big), \overrightarrow{x} \Big) \Big \rangle  \\
   & + \Big \langle f\Big(\Big(\theta_{\overline{j}, i}, \theta_j -\dfrac{\pi}{2}, \theta_i- \dfrac{\pi}{2}\Big), \overrightarrow{x} \Big) \Big \rangle \Big] \\
\end{split}
\end{equation}
Thus, the loss function $c(f(\theta, \overrightarrow{x}))$ curvature can be studied via second-order derivative of the loss. The chain rule is performed twice to obtain the Hessian matrix elements of \textit{c}.
\begin{equation}
	\begin{split}
	\partial \theta_i \partial \theta_j c(f(\theta, \overrightarrow{x}), y) & = \partial \theta_i \partial \theta_j f(\theta, \overrightarrow{x}) c^{'}(f(\theta, \overrightarrow{x})) + \partial \theta_j  \\
	& f(\theta, \overrightarrow{x}) \partial \theta_i  f(\theta, \overrightarrow{x}) c^{''}(f(\theta, \overrightarrow{x})) \\
\end{split}
\end{equation}

where $c^{'}$ and $c^{''}$ denote the first order and second order derivatives of the loss function, respectively with respect to the parameter ($\theta$). In this paper, we considered the cost (or loss) function $c(f(\theta, \overrightarrow{x}), y)=(y-f(\theta, \overrightarrow{x}))^2$, which will be minimized. On calculating the derivatives, the Eq. (6) becomes:
\begin{equation}
	\begin{split}
		\partial \theta_i \partial \theta_j c(f(\theta, \overrightarrow{x}), y) & = \partial \theta_i \partial \theta_j f(\theta, \overrightarrow{x}) (-2(y-f(\theta, \overrightarrow{x})))  \\
		& + \partial \theta_j  f(\theta, \overrightarrow{x})  \partial \theta_j f(\theta, \overrightarrow{x}). 2 \\
	\end{split}
\end{equation}
If the cost displays barren plateau, then its variance is vanishing exponentially as $Var[\partial_i f(\theta, \overrightarrow{x}))] \leq F(n)$,  where $F(n) \in \mathcal{O}(1/a^n)$, for $a>1$ \cite{19, 36, 37}. Then, the Hessian matrix elements are vanishing exponentially.

\section{Experiment Settings}
We have used the PyTorch library \cite{39}, and pennylane package \cite{38} for developing and training the variational quantum classifiers. The code is written in pennylane with a large number of parameters to expedite the experiments. The implementation is performed using PyTorch to accelerate the simulation using algebraic manipulation. All the quantum simulations are performed using python framework on the PennyLane platform for quantum differentiable
programming \cite{811}. The 2D and 3D graphs are plotted using a Plotly i.e. graphing library available in python.

\subsection{Warm-up Example}
\begin{figure}[!ht]
	\includegraphics[scale=0.55]{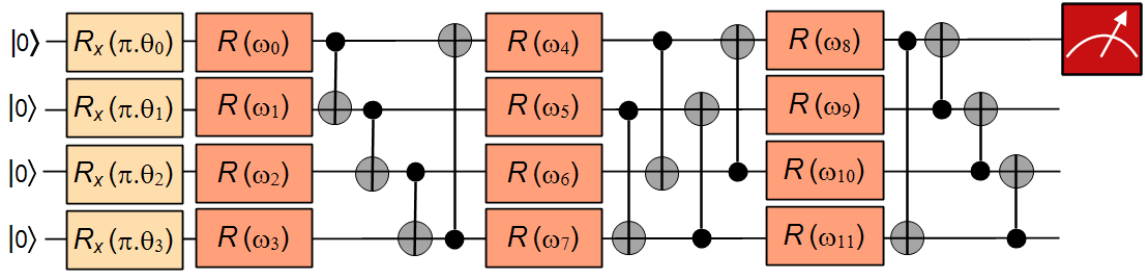}
	\caption
	{\small \textbf{Feature map and variational quantum circuits for the parity problem.} A four-qubit feature map consisting rotation around x-axis to prepare the initial state. It is followed by three layers of variational gates, where each layer consists of rotational gates with three trainable parameters ($R(\omega_j)=R(\phi,\theta,\omega)$) in each of the four qubits, followed by a set of CNOT gates. For two classes, the measurement is performed on one qubit, which is enough to have orthogonal measurements for the classes. }
\end{figure}

In this section, we characterized the loss landscape curvature of the four-bit parity problem with the Hessian. We begin exploring the concept of Hessian of loss functions of VQCs with a warp-up activity of solving a four-bit parity problem. Before illustrating the variational quantum classifier, the partial function is defined as a Boolean function whose output is 1 if and only if the input vector has an uneven number of ones. It is also called the XOR function of two inputs. The \textit{n}-bit parity function is given as \cite{40}
\begin{equation}
	f: v \in \{0, 1\}^{\otimes n} \rightarrow w \left\{
	\begin{array}{ll}
		1~ \text{if odd number of 1's in v}\\
		0~ \text{else} \\
	\end{array} \right \}
\end{equation}

The first step is to encode the input vectors into a quantum state. In a warm-up example, the inputs are 4-bit strings that are encoded into the state of qubits. The feature mapping and variational circuit of the four-bit parity problem are shown in Fig 2.  A single qubit rotation is defined as: 
\begin{equation}
R(\phi,\theta,\omega) = RZ(\omega)RY(\theta)RZ(\phi)= \begin{bmatrix}
	e^{-i(\phi+\omega)/2}\cos(\theta/2) & -e^{i(\phi-\omega)/2}\sin(\theta/2) \\
	e^{-i(\phi-\omega)/2}\sin(\theta/2) & e^{i(\phi+\omega)/2}\cos(\theta/2)
\end{bmatrix}.
\end{equation} The initial layer of $R_x$ gates prepares the initial state, which is also known as feature map circuit. Later, there are three layers of variational gates, where each layer consists of rotational gates with three trainable parameters in each of the four qubits, followed by a set of CNOT gates. It is to be noted that the combination of CNOT gates in each layer has a different structure.
The variation circuit consists of 36 parameterized gates and 12 non-parameterized gates.

The input data $\overrightarrow{v}$ is encoded into quantum state using feature map function as $ \psi: \overrightarrow{v} \rightarrow \ket{\psi( \overrightarrow{v})} \bra{\psi(\overrightarrow{v})}$. Initially, one qubit rotations are performed around \textit{x}-axis. It is followed by $\textit{U}_1$ single-qubit gate to apply a quantum phase to the qubit. Furthermore, a controlled-NOT multi-qubit gate is applied to flip the target qubit when the control qubit is in $\ket{1}$. The purpose of the feature map circuit is to map the classical input data into the quantum state.

The final \textit{n}-qubit feature quantum state becomes

\begin{equation}
	\ket{\psi({\overrightarrow{v}})}= U_{\psi(\overrightarrow{v})} \ket{0}^{\otimes n}
\end{equation}

\begin{figure}[!ht]
	\includegraphics[scale=0.55]{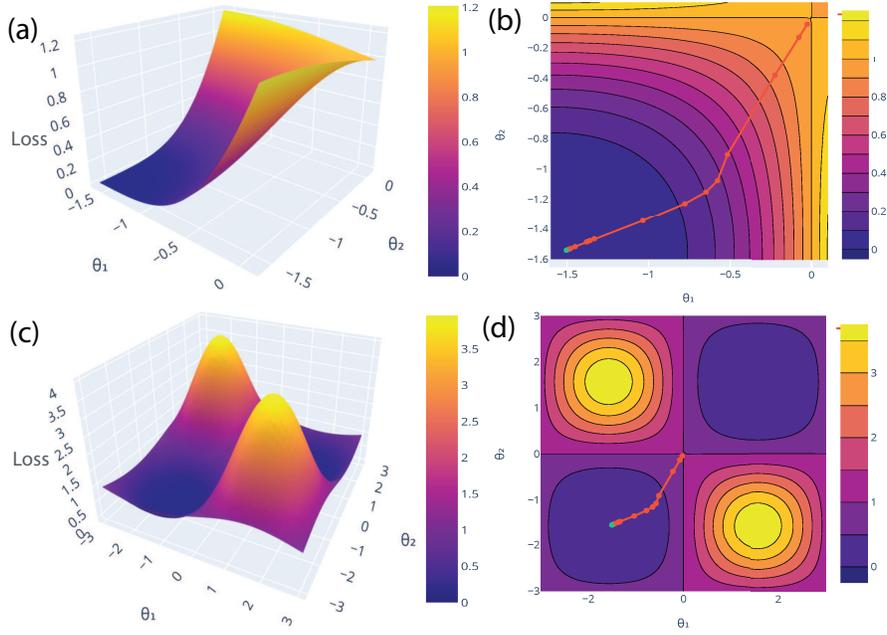}
	\caption{\textbf{Loss landscape of the parity problem for ($\theta_{3}$ and $\theta_{7}$)}. In fig (a, c), the loss landscape is visualized with a local loss function for two parameters $\theta_{3}$ and $\theta_{7}$, where rest of the parameters are set to the optimal values. Moreover, it cannot be visualize for more than 2 parameters for the 3D Loss,  because it cannot obtain the full range of loss between (0 and 1). In fig (b, d), the points in contour plots show the direction of improvement in the optimization process i.e. how the parameters are optimized during iterations. Fig (b) is just the zoom-in version of Fig (d) for better clarification. The green color point depicted the optimal value of $\theta_{3}$ and $\theta_{7}$.}
\end{figure}

Second, a short depth quantum circuit V($\theta$) is applied to the feature state. It depends upon the selection of parameterization for the gates and the number of \textit{d} layers. The classical optimizer handles the parameters during training to reduce the value of a loss function. Before returning a final classifier outcome, classical postprocessing is performed to the expectation value of the circuit. The aim is to determine the optimal classifying circuit $V(\theta)$ that separates the dataset with distinct labels. In variational quantum circuit, we used the  $R_y(\theta)$ and $R_z(\theta)$ parameterized gates that are applied  to rotate the qubits by angle ($\theta$) around \textit{y}-axis and \textit{z}-axis, respectively. CNOT gates follow it. The objective is to find a sequence of gates that forms a final state $\ket{\psi}_0$. A cost function ($C_f$) is defined as the square of trace distance ($D_t$) between final $\ket{\psi}_{0}$ and initial state $\ket{\psi}=V(\theta)^{\dagger}\ket{0}$, which is determined as
\begin{equation}
	C_f= \text{Tr}[O_f V(\theta) \ket{\psi}_0 \bra{\psi}_0 V(\theta)^\dagger]
\end{equation}

where $O_f=1-\ket{0}\bra{0}$. It is equivalent to $C_f=D_t(\ket{\psi}_0 \bra{\psi})^2$. As a warm-up to our study, the performance of a variational quantum classifier is tested for a parity problem. It is learnable in a quantum setting, a binary classification task $w=\{0, 1\}$ and its outcomes are measured on the computational basis. We selected the third qubit in the Pauli-\textit{Z} direction to perform the measurement and thresholding ($\varDelta$) the expectation value  $\langle Z \rangle  \leq \varDelta$ ($\langle Z \rangle  > \varDelta$) to classify the input vectors in to one of the labels \textit{w}=0 (\textit{w}=1), respectively. We utilized the gradient descent optimizer (GD) to iterate a parameter update based on the gradient of the loss function. It is used to minimize an objective function to its local minimum by adjusting the parameters repeatedly. The partial derivatives of the cost function are determined with respect to each parameter and store the outcome in a gradient. A step of the GD optimizer determines the new values via the rule $\theta^{(t+1)} = \theta^{(t)} - \eta \nabla f(\theta^{(t)})$, where  $\eta$ is a user-defined hyperparameter relating to step size.  We now consider the concept of Hessian that how it helps to realize the loss landscape (or curvature information) for the parity problem. Firstly, the parameters are initialized randomly and determine the set of parameters that can produce the target state. The optimization problem is translated into the loss function minimization.

\begin{figure}[!ht]
	\includegraphics[scale=0.55]{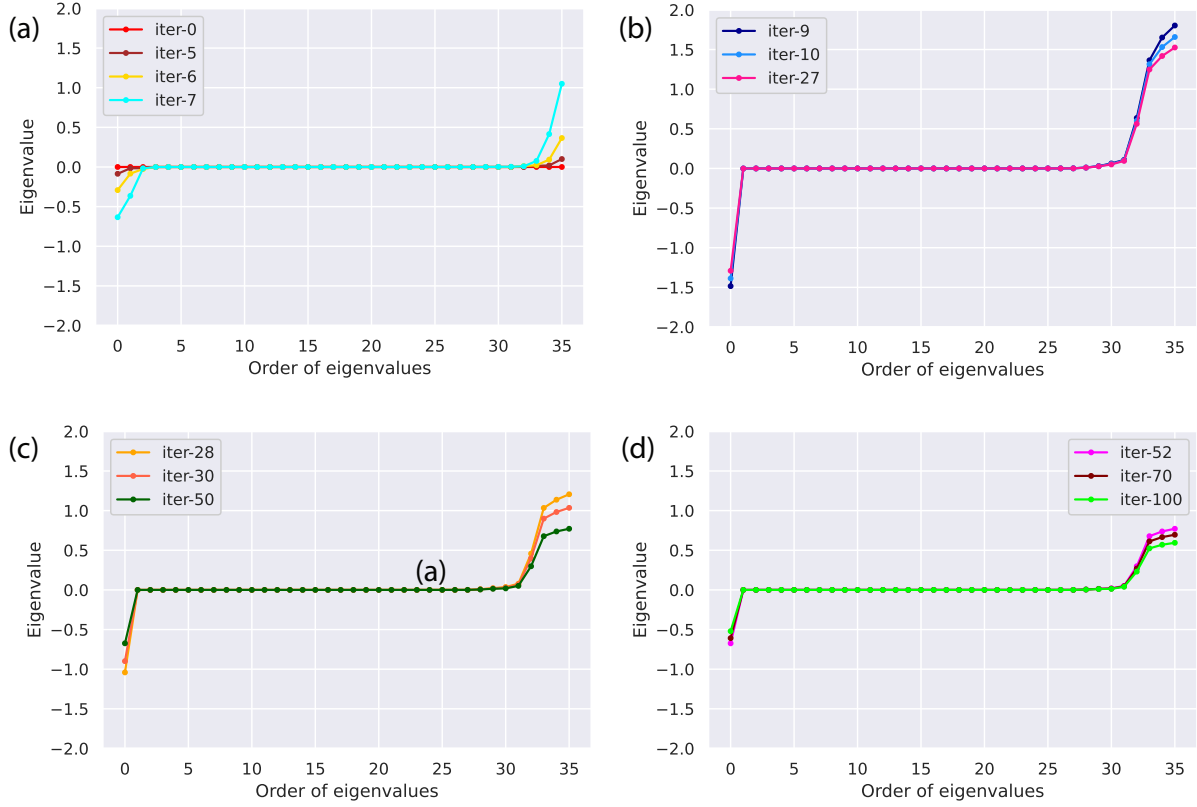}
	\caption{\textbf{The evolution of behavior of the eigenvalues of
			the Hessian during training for the parity problem}. Initially, it shows the mixture of non-negative and negative values. Finally, a well converged loss is observed where the most of the Hessian's eigenvalues are non-negative.  During the training,  we separate the curves for better clear view as the difference
		between smallest and biggest eigenvalues is changing for different epochs. } 
\end{figure}

 It is presented as a function of $\theta_3$ and $\theta_7$, and rest of the parameters are set to the optimal values. Fig 3 shows the loss landscape of a parity problem with a local loss function of parameters, where ($\theta_1=\theta_3$ and $\theta_2=\theta_7$). We started with some random initialization. The contour plots Fig 3 (b) and (d) show the direction of improvement in the optimization process that how the parameters  ($\theta_3$ and $\theta_7$) are optimized during 100 iterations. The green color point denotes the optimal value. The zoom-out and zoom-in versions of contour plots are shown in Figs (b) and (d) for a better view of optimal values. 

 The minimum value of a loss can be recovered with the gradient descent optimizer due to the point of local convexity in the landscape. Fig 4 depicts the behavior of eigenvalues of the Hessian for parity problem during training. During the optimization process, the Hessian matrix is calculated on all the trainable parameters and eigenvalues are recorded after each iteration. In Fig 4, the eigenvalues are plotted for some specific iterations in ascending order to observe how the behaviour of the eigenvalues is changing with the trainable parameters being updated during the progression of an optimization process. Fig 4 (a-d) shows the variations between the minimum and maximum eigenvalues at each iteration. The distribution of eigenvalues for the randomly initialized quantum circuit shows the mixture of negative and non-negative values close to zero at iterations (0-7). Fig 4 (b-c) shows some of the eigenvalues are positive, some are negative, and the bulk of them is zero. Fig 4 (d) shows a loss for the well converged variational circuit where we left with a single negative eigenvalue of the Hessian and rest all are non-negative (at 100th iteration). In fact, the zero gradient ensures that it is a global minimum. Moreover,  the zero eigenvalues correlate to directions where variations in parameters do not alter the curvature information.  Thus, Hessian's behavior of the eigenvalues helps to obtain the minimum and maximum stability in the directions of loss landscape.

\subsection{Classification of Diabetes}
Let us consider the case of the classical diabetes dataset for the classification of diabetes in supervised learning. In this section, we analyzed the loss landscape through the Hessian for diabetes dataset and investigated how VQC will perform to predict diabetes. The diabetes dataset can be download from UCI machine learning repository \cite{44}. It consists of 8 input features (age, glucose, insulin, pregnancies, body mass index (BMI), skin thickness, diabetes pedigree function, and blood pressure) and one binary output feature.

 \begin{figure}[!ht]
	\includegraphics[scale=0.5]{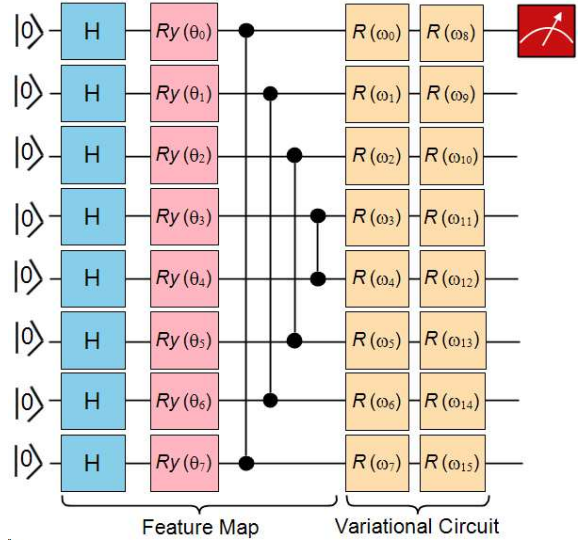}
	\caption{\label{fig:epsart}\textbf{Variational quantum circuit}. A general 8-qubit variational quantum circuit is constructed for diabetes dataset with a local cost function. In feature map,  Hadamard gates, rotation around y-axis and control-Z entangling gates are applied. The variational part of circuit consisting parameterized rotations $R(\omega_j)=R(\varphi_1, \varphi_2, \varphi_3)$ applied on each qubit. }
\end{figure}

\begin{figure}[!ht]
	\includegraphics[scale=0.5]{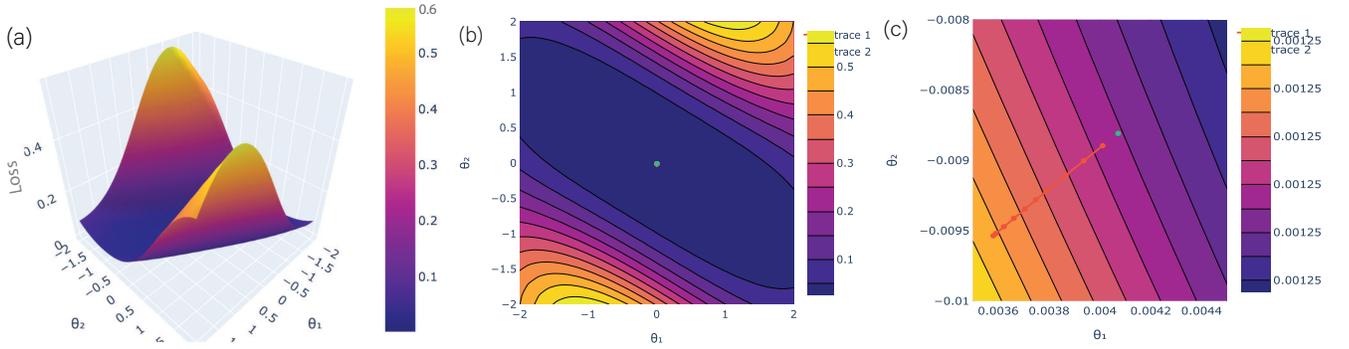}
	\caption{\textbf{Loss landscape of the diabetes dataset for ($\theta_{0}$ and $\theta_{24}$)}. (a) The loss landscape is demonstrated with a local loss function for two qubits $\theta_{0}$ and $\theta_{24}$. Fig (b-c) The direction of improvement in the optimization process is shown using contour plots i.e. how the parameters are optimized during iterations.}
\end{figure}

\begin{figure}[!ht]
	\includegraphics[scale=0.5]{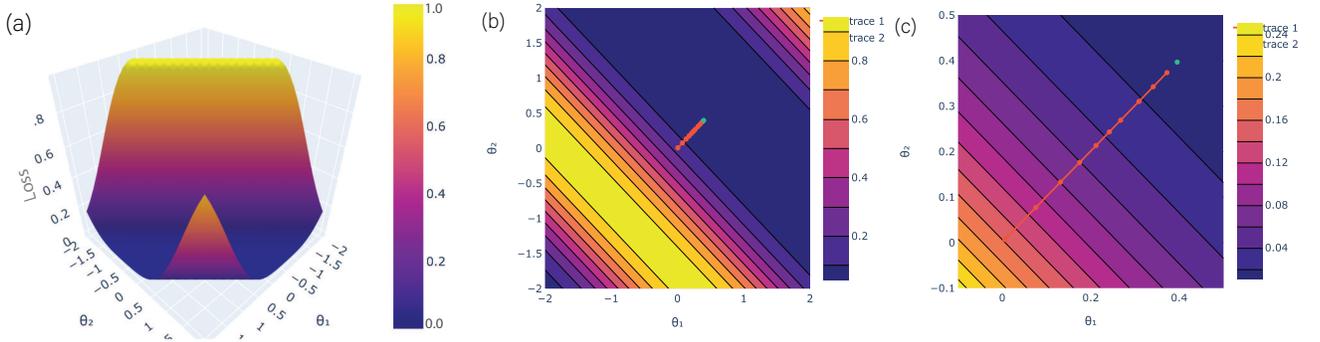}
	\caption{\textbf{Loss landscape of the diabetes dataset for ($\theta_{1}$ and $\theta_{25}$)}. (a) The loss landscape of the diabetes dataset is visualized with a local loss function for two parameters $\theta_{1}$ and $\theta_{25}$. Fig (b-c) give a clear view of the optimal values during iterations.}
\end{figure}

Firstly, the 8 input features are encoded into the state of qubits by a quantum feature map. Consider a classical dataset $D = \{(x^n, y^n)\}^{S}_{n=1}$ for binary classification, where $y^n\in \{0, 1\}$ i.e. 0 for no diabetes and 1 is for diabetes. Each segment of classical data is encoded into an amplitude of a qubit using single-qubit rotations. Afterward, a variational quantum circuit $V(\theta)$ is applied to the feature quantum state for training and classification of diabetes. The 8-qubit variational quantum circuit with a local cost function is constructed, as shown in Fig. 5. The feature map consists of Hadamard gates, rotation around y-axis and control-Z entangling gates. It is followed by a variational part of the circuit containing single-qubit rotations $R(\omega_j)=R(\varphi_1, \varphi_2, \varphi_3)$ on each qubit.

\begin{figure}[!ht]
	\begin{subfigure}{0.5\textwidth}
		\centering
		\includegraphics[scale=0.45]{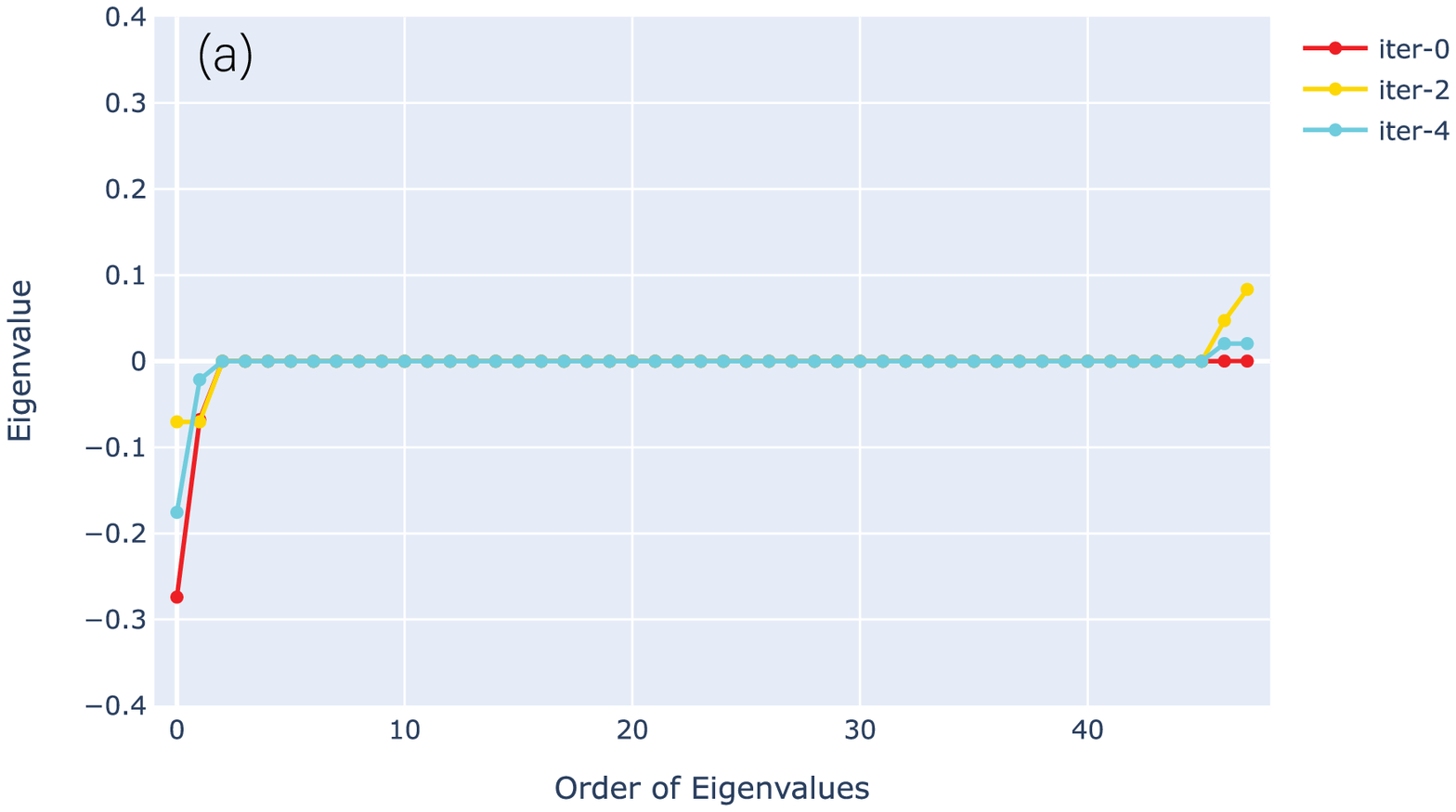}
	\end{subfigure}%
	\begin{subfigure}{0.5\textwidth}
		\centering
		\includegraphics[scale=0.45]{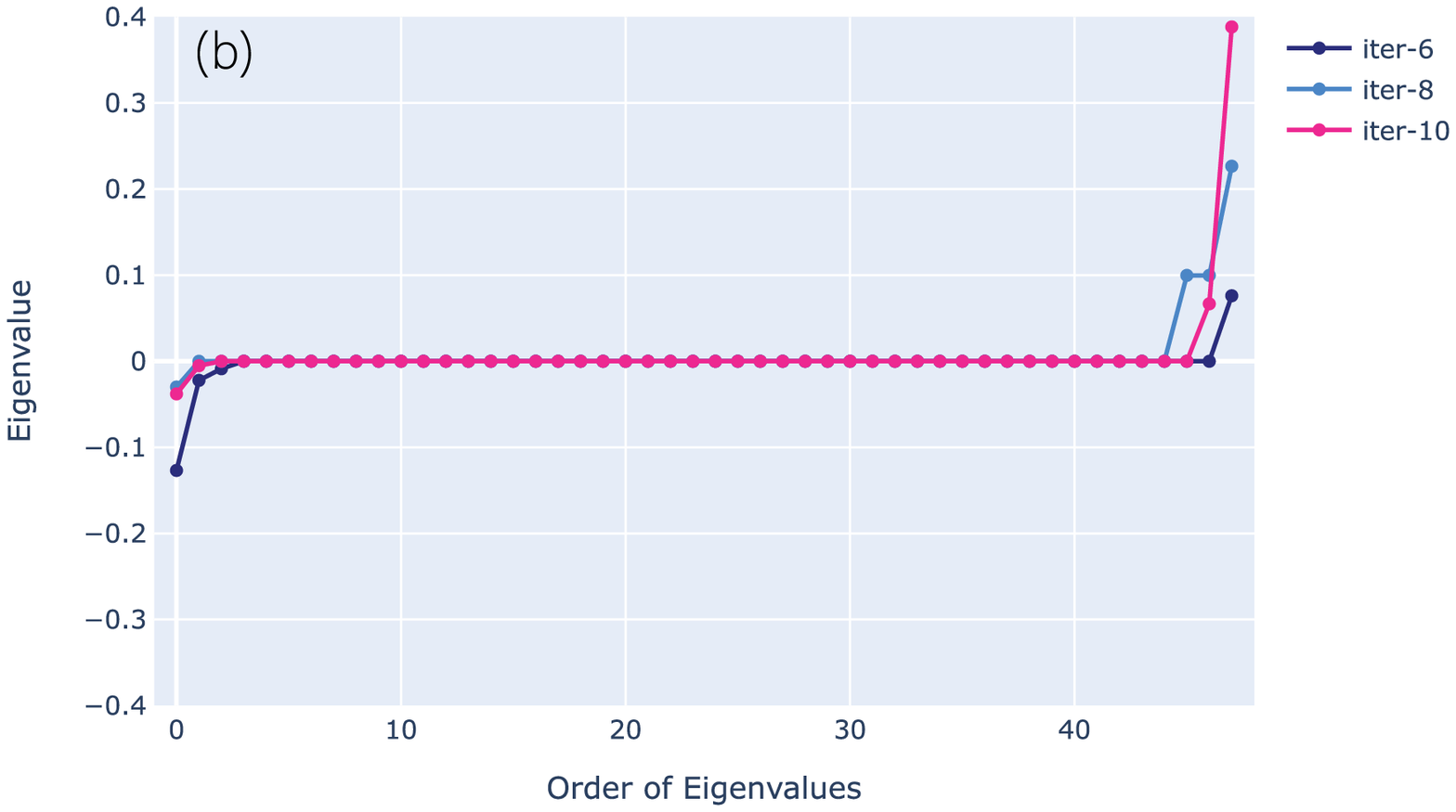}
	\end{subfigure}%
	\vspace{0.001cm}
	\begin{subfigure}{0.5\textwidth}
		\centering
		\includegraphics[scale=0.45]{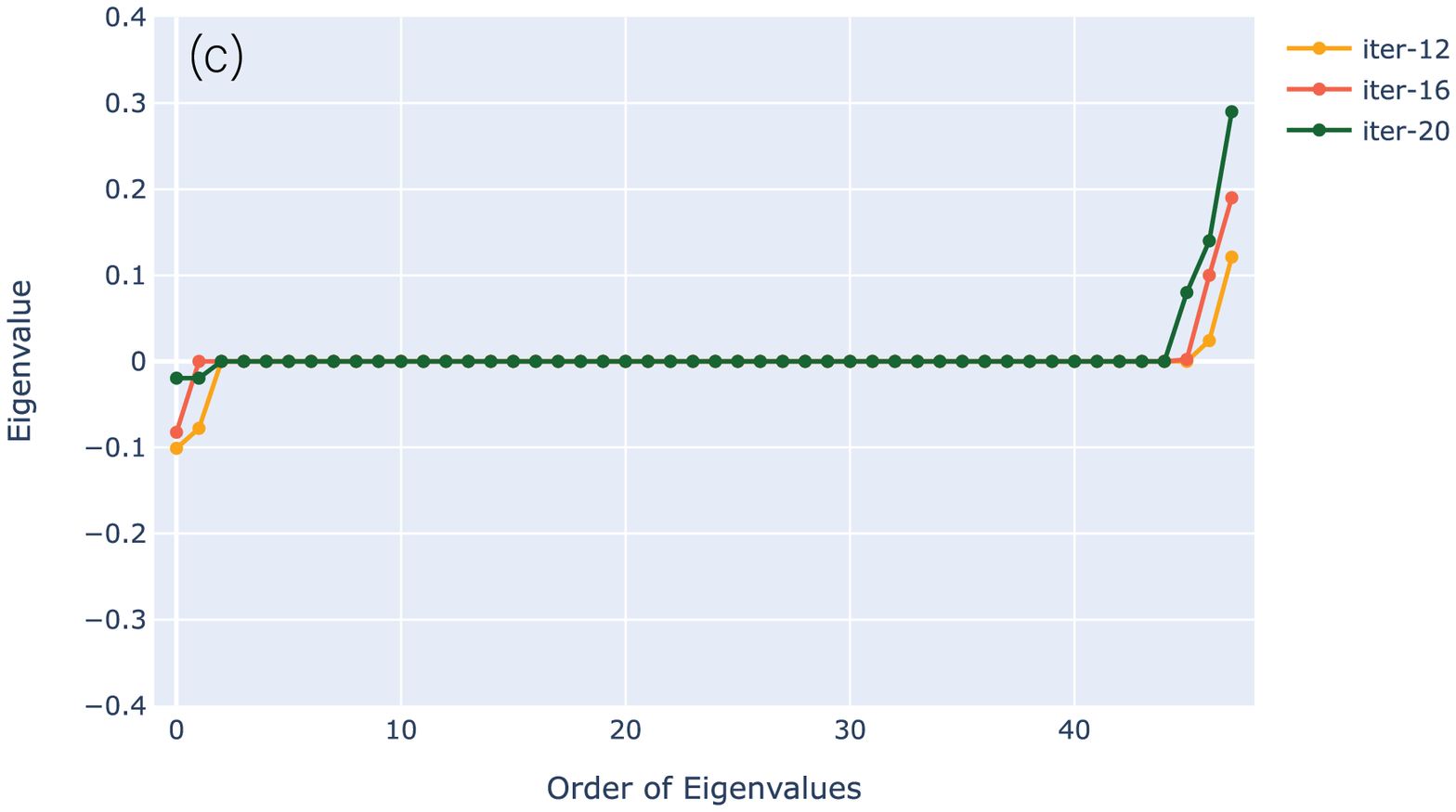}
	\end{subfigure}%
	\begin{subfigure}{0.5\textwidth}
		\centering
		\includegraphics[scale=0.45]{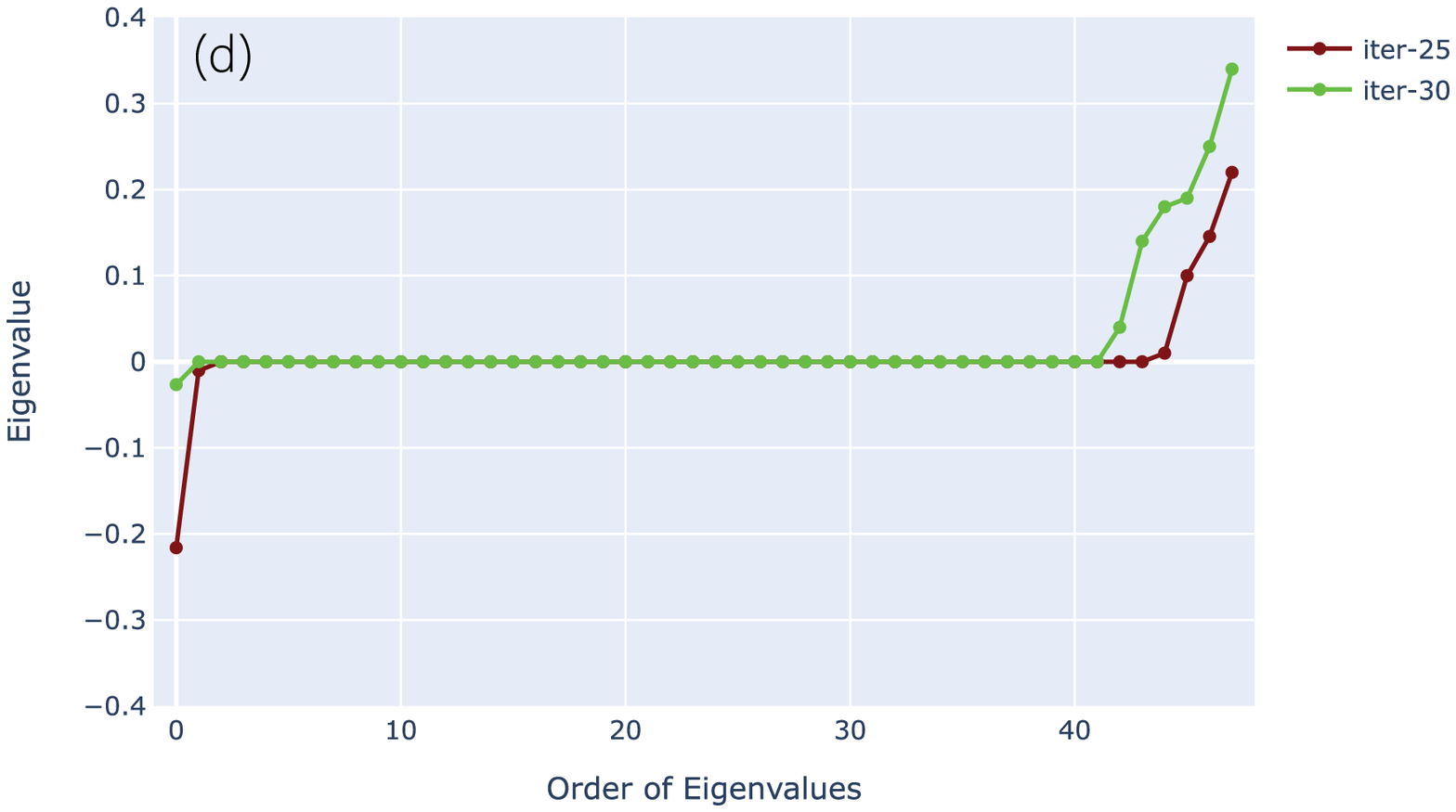}
	\end{subfigure}%
	\caption{\textbf{The evolution of behaviour of the eigenvalues of
		the Hessian during training for the diabetes dataset.} It shows the eigenvalues evolution during the training for different epochs. Initially, the Hessian's eigenvalues consisting of negative and positive values equally, and many of them are zero. After the convergence, it shows maximum positive eigenvalues.}
\end{figure}

The feature map and variational quantum circuits are repeated two times. It is constructed simple enough so that it can be executed on real quantum systems and complex to separate the input data after mapping. For our experiments, the feature map and variational quantum circuits were constructed to a fixed number of depth and qubits. To evaluate the most likely state of an outcome qubit, the quantum circuit can be determined for several iterations by considering a similar input to calculate the probability distribution among the basis states. We used the gradient descent optimization method to find the parameters which can determine the probabilities closest to reality. The measurement is performed on a qubit by employing the Pauli operators in a specific direction. The outcome qubit discovers the predicted value of an input, i.e., the class label values allocated. The loss landscape of VQC for a diabetes dataset with a loss function of parameters ($\theta_0$) and ($\theta_{24}$) is visualized in Fig 6, where $\theta_1$ and $\theta_2$ are set to $\theta_0$ and $\theta_{24}$, respectively. Here, the other parameters are set to the optimal value after each iteration. In each iteration, the workflow of the optimization process consists of three steps feature map, variational circuit and observation. These are performed on quantum circuits for each of the samples on the dataset and the loss is calculated by taking the distance of prediction from the label for each sample and then averaging them. It is further used in the classical optimization process of the trainable parameters of the variational circuit and finally the trainable parameters are updated as the final step of each iteration. 
During the observation purposes, the hessian matrix  is calculated on the trainable parameters in each iteration and then the eigenvalues of the hessian matrix are recorded only  for the data visualization and need not to be incorporated while being implemented on practical purpose. The input vectors are normalized to lie in $[-\pi, \pi]$ and plotted the loss landscape for two qubits in a range. If  we try to plot the loss landscape for more than two parameters, then it cannot obtain the full range between (0 and 1). The contour plots (b) and (c) show the evolution of optimized parameters during 30 iterations in Fig 7.  Furthermore, the loss landscape with a loss function of parameters ($\theta_1$) and ($\theta_{25}$) has been analyzed in Fig 7. It has been observed from the landscape that optimum cost is single with good local minima. It shows the prediction map of VQC with Z-measurement for the diabetes dataset.

The distribution of Hessian's eigenvalues is determined over the training process of a variational circuit  to locate one of the minima.  It is used to investigate whether a particular stationary point is a saddle point or not. At the beginning of the training, the gradient descent method is struggling to break the symmetry. Due to the small gradients, it faces a problem in training small quantum circuits. Fig 8 (a) shows a distribution of the Hessian's eigenvalues consisting of equally possible negative and positive values, and most of them are zero for iterations (0, 2, and 4). We observed that the negative eigenvalues gradually started to disappear with the increase in number of iterations, as shown in Fig 8 (c). After the convergence at 30th iteration, a single negative eigenvalue is left and rest all became non-negative, in Fig 8 (d). The bulk of zero eigenvalues shows a flat direction of the $f(\theta,  \overrightarrow{x})$ in the loss landscape, where any alterations in circuit parameters do not disturb the loss landscape.  The positive semi-definite behavior of the Hessian's eigenvalues signifies a very steady result.  Although, it is not practical to visualize the loss landscape of VQC in 3-dimensional due to the problem of fixing the other parameters. Nevertheless, it is feasible to visualize the loss landscape of variational quantum classifiers through the lens of the Hessian. 

\section{Convergence via Adaptive Hessian Learning Rate}

In this section, we show how the Hessian can adjust adaptively to the learning rate (LR) for each parameter. Adaptive learning rate has been a popular practice in classical machine learning. In any gradient based optimization method, a very large learning rate can cause an overshoot in cost while coming close to smaller gradients, whereas very small LR will cause an extremely slow approach towards the lowest point in loss landscape. Therefore, a smart trade off between both leads to the requirement of tuning the Learning rate properly. It is not as simple as it looks. We used an adaptive learning rate, which initially starts with the higher values of learning rate and gradually reduces it with reducing gradient values.

Adaptive Hessian learning rate (A-HLR) is something similar to the previous concept with an add on, specific to the quantum machine learning model's loss landscape analysis. Here, instead of using continuous variation of learning rates, we consider a set of discrete values of learning rates. We begin with the largest LR and gradient descent optimization process. After each step, we compare the updated cost with the previous cost. After few consecutive steps, if the difference between updated  and previous cost is below a threshold level, then LR is set to the next smaller discrete value in the set of LR's. The complete process is repeated until we reach the lowest LR in the set of discrete learning rates. Since the loss landscape of quantum machine learning models is quite different from the classical machine learning models. There can be observed a specific repetitive manner of landscape unlike classical model, which leads to the higher probability of getting stuck in local minima.

In case, if the execution reached lowest LR among the set of discrete learning rates, there can be two possible cases (i) either the optimization stuck in a local minima, we need to get out of this, (ii)  or the optimization problem reached the global minima. To tackle these scenarios, we considered the concept of Hessian matrix and used it with a set of decreasing learning rates. The implementation is started with the largest learning rate, for each of values of the learning rate in the set and the process of optimization continues until the difference of loss values in two consecutive iterations goes below tolerance. Once it occurs,  then the next learning rate is considered from the predefined set and the process is repeated until it reaches the least learning rate value.  If the count of negative eigenvalues is lesser than a threshold value, then the solution obtained is a global minima. In case, if the count is greater, then it got stuck in local minima. It is to be noted that the threshold value of negative eigenvalue count is a hyperparameter, which has to be tuned properly depending on the type of data in a given dataset. (Practically it should be zero, but in reality its not always possible to reach that point). If the model is stuck at local minima, i.e. the learning rate has already reached the lowest of its all possible given values. Then, the optimization will be again started with the higher learning rate with an objective that the cost will overshoot and come out of the local minima. The process is repeated until the LR comes to its lowest value. Finally, it is  evaluated whether the optimization has reached global minima or not using the Hessian matrix.

\begin{figure}[!ht]
	\includegraphics[scale=0.5]{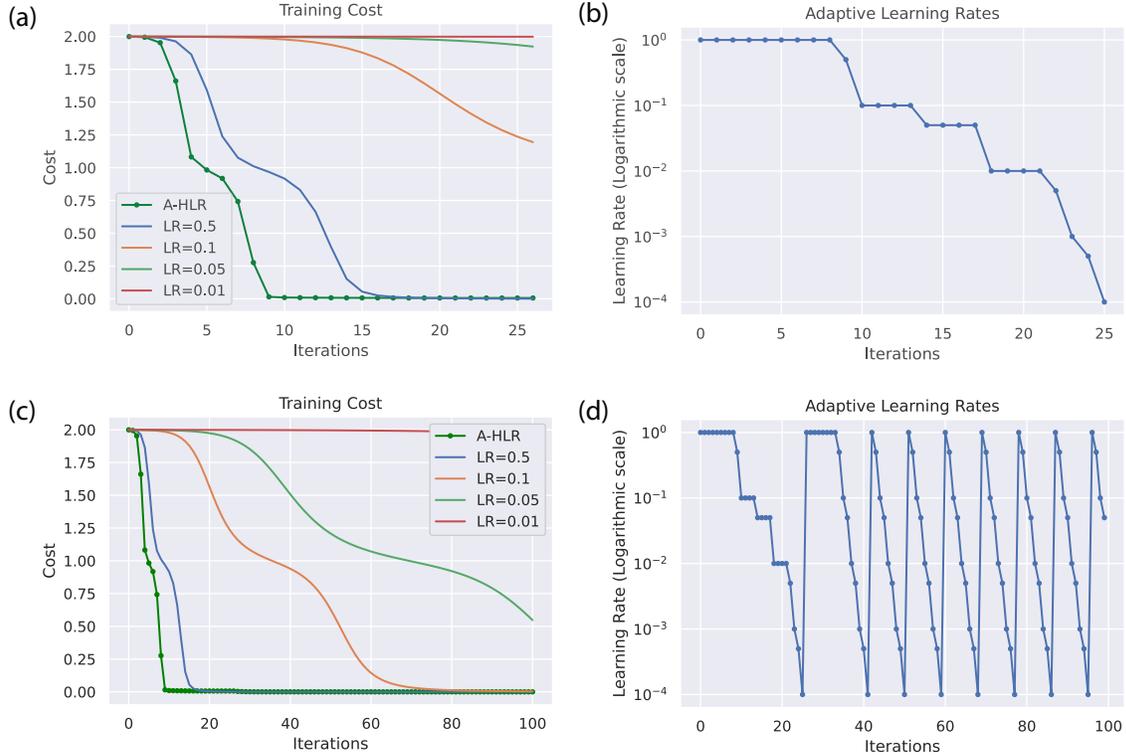}
	\caption{\label{fig:epsart}\textbf{Training cost of four bit parity dataset with adaptive Hessian learning rate (A-HLR).} In fig (a, c), the comparison is shown between the adaptive Hessian learning rate and gradient descent method with constant learning rates. A-HLR shown faster convergence than gradient descent with fixed learning rates.  It has been shown the stable and efficient convergence of the cost function for parity dataset during training. Fig (b, d) shows how A-HLR evolves during training of 25 and 100 iterations. It depicts how to overshoot the local minima of the cost function using adaptive Hessian lower to higher learning rates.}
\end{figure}

Fig 9 (a) shows the comparison that how the cost is evolving throughout the optimization process with an adaptive Hessian learning rate (A-HLR) method and  gradient descent method with constant learning rates. It has been determined that A-HLR converged very well within 25 iterations. The gradient descent method with LR=0.5 learning rate also converges, but not as quick as with A-HLR. Fig 9 (b) shows how the value of adaptive learning rate evolves on using A-HLR during first 25 iterations and depicts how it overshoots the local minima of the cost function using lowest to highest LRs.
Fig 9 (c) shows the comparison between A-HLR and gradient descent methods with different learning rates for 100 iterations. Fig 9 (d) depicts how the value of adaptive learning rate evolves on using A-HLR during the first 100 iterations. A-HLR fits the local shape of gradient to the loss landscape very well, provides a
faster convergence than gradient descent method with constant learning rates. It permits one to select a descent direction for faster convergence during the training
of variational circuits. Therefore, the local traps in the loss landscape can be avoided by using the adaptive hessian learning rate approach.

\section{Conclusion}
In this paper, the curvature information of the loss landscape of variational quantum classifiers has been visualized via the lens of Hessian eigenvalues. We developed a simple theoretical quantum model of Hessians and gradients, as justified by datasets for numerical justifications on VQCs. The parity function problem is considered as a warm study to show the behavior of Hessian's eigenvalues. It has observed that VQC has an exceptional ability to generalize small datasets. Furthermore, we visualized the cost function landscape of VQC designed for the diabetes dataset. It converges efficiently for data-driven problems. We identified some differences in the convergence with adaptive Hessian learning rate and gradient descent method using fixed learning rate. It has been observed that adaptive Hessian learning rate helps to overshoot the cost if it gets fall into local minima and converge quickly. It seems beneficial to study the local curvature information of VQC through the Hessian. The integration of gradient-based methods and Noisy Intermediate-Scale Quantum (NISQ) devices is still a young area and potentially has a lot more to offer. In the future, this work will open up new avenues of research in solving classical and quantum optimization problems and framework design. It will help the research communities to accelerate the analysis of variational quantum algorithms based on Hessian.

\section*{Data Availability}
The data that support this study are available from the corresponding author upon
reasonable request.

\section*{Author Contributions}
All authors have contributed equally to this work.

\section*{Additional information}
\textbf{Competing interests:} The authors declare no competing interests.

\end{document}